# Comparison of RABITT and FROG measurements in the temporal characterization of attosecond pulse trains

Kyung Taec Kim,[1] Kyung Sik Kang,[1] Mi Na Park,[1] Tayyab Imran,[1] G. Umesh,[2] and Chang Hee Nam[1]

[1]*Department of Physics and Coherent X-ray Research Center,*
*Korea Advanced Institute of Science and Technology (KAIST), Daejeon 305-701, Korea*
[2]*Physics Department, National Institute of Technology Karnataka,*
*Surathkal, Mangalore 575025, India*

The attosecond high harmonic pulses obtained from a long Ar-filled gas cell were characterized by two techniques - the reconstruction of attosecond beating by interference of two-photon transition (RABITT) and frequency-resolved optical gating (FROG) methods. The pulse durations obtained by RABITT and FROG methods agreed within 10 %.

The temporal characterization of attosecond pulses obtain from high-order harmonics is an important issue for applications of the attosecond pulses. The first temporal characterization was done using a method called the reconstruction of attosecond beating by interference of two-photon transitions (RABITT) by Paul *et al.* [1]. The RABITT method is a cross correlation technique based on the photoionization process realized by applying harmonic and IR femtosecond laser pulses with time delay. In addition to the photoelectron peaks generated by harmonic photons the sideband photoelectrons are generated between two adjacent harmonic photoelectron peaks. With the time delay these sidebands are modulated, providing the information on the spectral phase difference between adjacent harmonics. However, this method provides only averaged temporal characteristics of pulses existing in the attosecond pulse train.

For full temporal reconstruction of attosecond pulses, more rigorous techniques are needed. An improved technique, called the frequency-resolved optical gating (FROG) for complete reconstruction of attosecond bursts, was proposed by Mairesse [2]. The conventional FROG technique, widely used for the characterization of femtosecond lasers, is applied to the reconstruction of the temporal structure of harmonic pulses from the photoionization process in strong laser field. While the spectral shape of the harmonics is assumed to be equally spaced with zero bandwidth in the RABITT method, no such assumption on the spectral shape of attosecond pulses is necessary for the FROG method. Consequently, the complete temporal information on attosecond harmonic pulses can be found using this reconstruction method, applicable both for attosecond pulse train and single attosecond pulse.

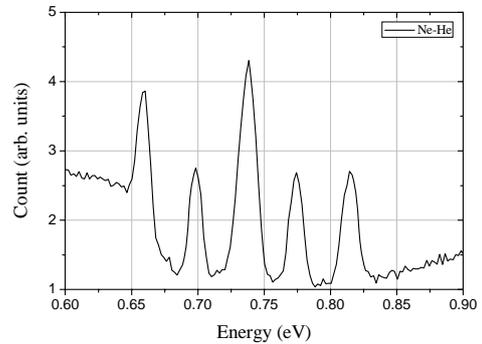

**Fig**. 1 Energy resolution measurement of the TOF spectrometer. The photoelectron energy of He atoms ionized by Ne III spectral lines was measured with TOF. The energy resolution of 11 meV was obtained at the electron energy of 0.7 eV.

Here we have carried out a direct comparison of RABITT and FROG measurements in order to confirm the validity of the RABITT measurement. We experimentally demonstrated the FROG technique for the temporal characterization of attosecond harmonic pulses, especially attosecond pulse trains [3]. As in the case of RABITT, the photoionization process with harmonic and IR laser pulses was used in the FROG measurement. The FROG measurement, however, required much more extensive data (~10 times) with good energy resolution. For this purpose we recently improved further the energy resolution of our time-of-flight (TOF) electron spectrometer, as presented in Fig. 1, together with good energy calibration. This enabled us to make more accurate and reliable FROG measurements for the temporal characterization of attosecond harmonic pulse trains, allowing a quantitative comparison between the two measurements.

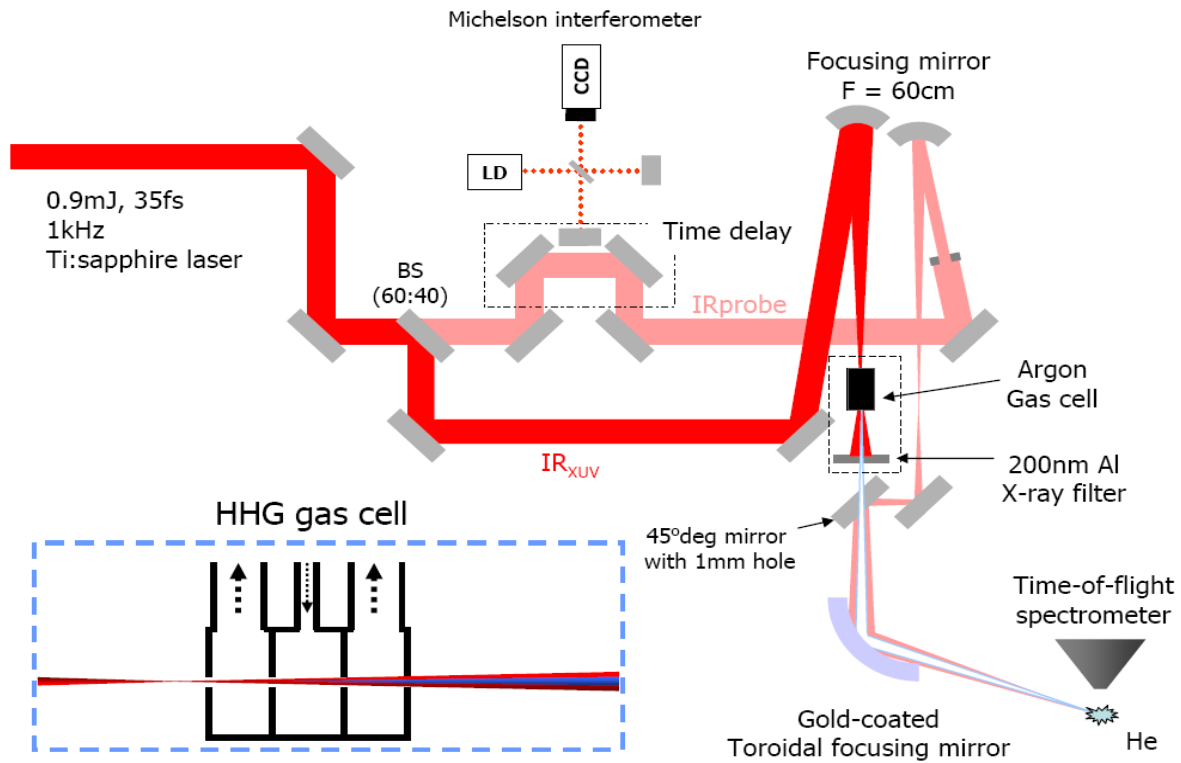

**Fig**. 2 Experimental setup.

For reliable FROG measurements a series of experiments were carried out. Firstly, the energy resolution of our TOF electron spectrometer was measured using the photoionization of helium atoms by neon III spectral lines, as in Fig. 1. The energy resolution is seen to be 11 meV at the electron energy of 0.7 eV. Secondly, the photoelectron spectra were recorded over a full range of the time delay scan between harmonic and IR femtosecond laser pulses with delay time steps of 160 as. In the experiments, 35-fs, 822-nm Ti:sapphire laser operating at 1-kHz was used for harmonic generation and probe. Harmonics were generated in a 12-mm Ar-filled gas cell placed 4-mm behind the focus. For an accurate time-delay control the mirror movement by PZT was monitored by an interferometer, as shown in the experimental setup in Fig. 2. The laser intensity used for the harmonic generation was $2 \times 10^{14}$ W/cm$^2$ and that for the probe pulse was $1 \times 10^{12}$ W/cm$^2$.

The harmonic pulse train was reconstructed using the principal component generalized projection (PCGP) algorithm of FROG [4]. For the FROG analysis the time delay scan was performed so as to sufficiently cover the overlap between the harmonic and probe IR laser pulses. The photoelectron spectra obtained with the harmonics from the 40-Torr Ar gas cell is presented in Fig. 3 (a) and the reconstructed spectra obtained using PCGP algorithm is shown in Fig. 3 (b). The reconstructed temporal profile in Fig. 3 (c) shows the temporal profile of the attosecond pulse train with a pulse duration of 6 fs. It also reveals that the attosecond pulse in the center is shortest, agreeing with a high harmonic simulation in a long gas jet [5].

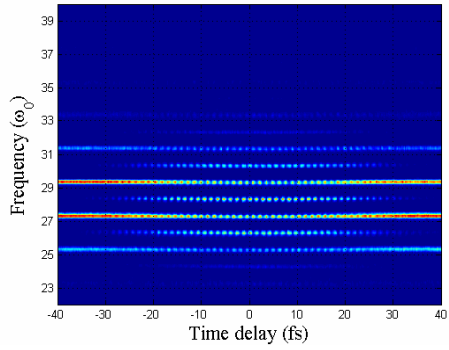

(a)

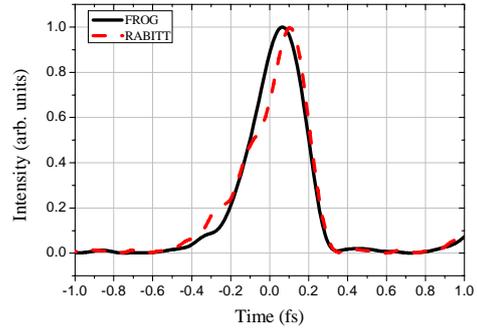

(a)

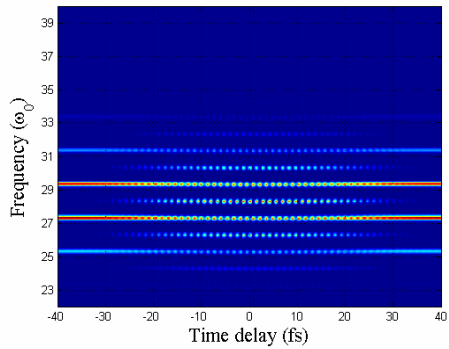

(b)

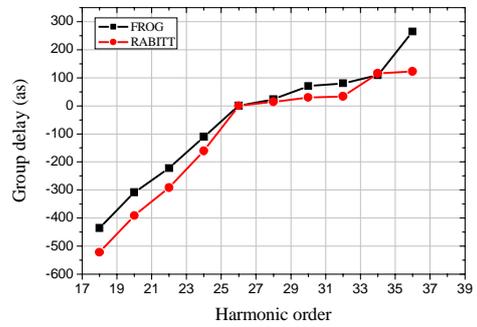

(b)

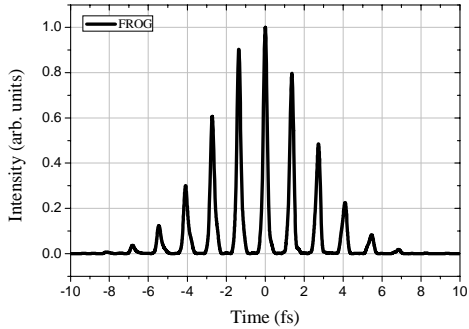

(c)

**Fig**. 3 (a) Photoelectron spectra obtained with a full range scan of time delay between harmonic and IR femtosecond laser pulses, (b) reconstructed photoelectron spectra, and (c) reconstructed attosecond pulse train with the PCGP algorithm of FROG. The harmonics were generated from the Ar gas cell at a pressure of 40 Torr.

Direct comparison of the FROG and RABITT measurements was carried out using same photoelectron spectra. Two cases of these measurements - one in low Ar pressure and the other in high Ar pressure - were performed. In the case of

**Fig**. 4 Comparison of the FROG and RABITT measurements for the case of 15-torr Ar. (a) Temporal profiles and (b) group delays are shown. The measured durations are 296 as and 302 as for the RABITT and FROG measurements, respectively, while the transform-limited pulse duration is 144 as.

15-torr Ar gas medium, the measured pulse durations were 296 as and 302 as for the RABITT and FROG measurements, respectively. In the case of FROG the strongest attosecond pulse at the center was used. Figure 4 shows the temporal profiles and group delays of the RABITT and FROG measurements. As the transform-limited pulse duration was 144 as, the results show clearly the effect of strong attosecond chirp, as evident in the group delay data. In the case of 40-torr Ar target, the pulse durations measured were 226 as and 232 as for the RABITT and FROG measurements, respectively. Figure 5 shows the temporal profiles and group delays. The measured duration shows good chirp compensation in the harmonic generation medium, as the transform-limited duration was 211 as. These measurements show that the two methods agree in the pulse duration within 10 %.

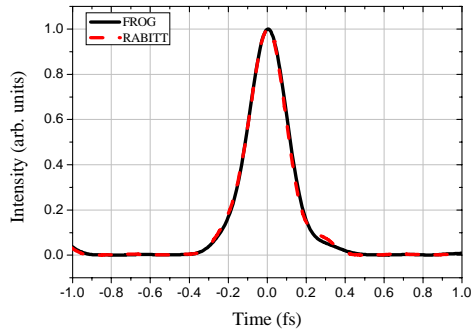

(a)

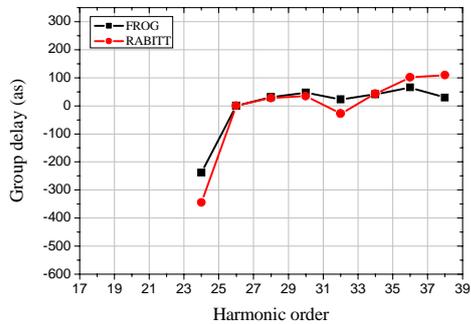

(b)

**Fig**. 5 Comparison of the FROG and RABITT measurements for the case of 40-torr Ar. (a) Temporal profiles and (b) group delays are shown. The measured durations are 226 as and 232 as for the RABITT and FROG measurements, respectively, while the transform-limited pulse duration is 211 as.

The direct comparison shown above clearly indicates that the pulse duration measured by the RABITT method is sufficiently accurate. In addition, we show, in Fig. 6, that the harmonic blueshift can be fitted with a linear curve. The blueshift causes slight shortening of the period in the attosecond pulse train. This does not create any serious problems in the application of the RABITT method.

In summary, with regard to the measurement of pulse duration, the RABITT measurements showed good agreement with the FROG results. Though the RABITT method cannot yield the temporal characteristics of individual pulses in an attosecond pulse train, it is useful for obtaining the information on averaged temporal characteristics, and, especially, for monitoring the change of pulse duration. It is clear, therefore, that the RABITT technique is sufficiently accurate for the measurement of attosecond pulse durations.

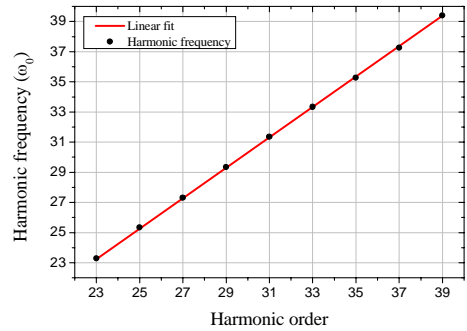

**Fig**. 6 Harmonic frequency fitted with a linear curve, for the case of 40-torr Ar, with $\omega_q = 1.01\omega_0$.